# Electromagnetic radiation and resonance phenomena in quantum mechanics

V. A. Golovko

Moscow Polytechnic University, Bolshaya Semenovskaya 38, Moscow 107023, Russia

Abstract

It is demonstrated that, if one remains in the framework of quantum mechanics taken alone, stationary states (energy eigenstates) are in no way singled out with respect to nonstationary ones, and moreover the stationary states would be difficult if possible to realize in practice. Owing to the nonstationary states any quantum system can absorb or emit energy in arbitrary continuous amounts. The peculiarity of the stationary states appears only if electromagnetic radiation that must always accompany nonstationary processes in real systems is taken into account. On the other hand, when the quantum system absorbs or emits energy in the form of a wave the determining role is played by resonance interaction of the system with the wave. Here again the stationary states manifest themselves. These facts and influence of the resonator upon the incident wave enable one to explain all effects ascribed to manifestation of the corpuscular properties of light (the photoelectric effect, the Compton effect etc.) solely on the basis of the wave concept of light.

# 1. Introduction

Amongst elementary particles, the most mysterious particle is the particle under number one, namely, the photon. As to other elementary particles, there exist judgments concerning their structure or at least one could estimate their size. In the matter of the photon, not only do any conjectures as to its structure lack but even there are no reasonable guesses about its dimensions. In this connection it should be remarked that the dual (particle-like and wave-like) properties of an object require two parameters with a dimension of length. For such particles as an electron or a proton, the size and de Broglie wavelength are quite different things, the latter being dependent in addition on the velocity even in the nonrelativistic case. If the object is characterized by only one parameter of length dimension, the object is either a particle or wave. The photon possesses only one characteristic length, the wavelength. Particularly striking are the properties of the photon once this notion is applied to radio waves that represent a special case of electromagnetic radiation. What could be said, for example, about the size of the photon (regarded as an elementary particle) in the case where the wavelength of a radio wave amounts to several kilometres? Once the photons of an electromagnetic wave are separate particles with different energies, if the wave is not monochromatic, a question may also be raised as to what forces keep the photons from diverging when the wave emitted by a remote galaxy travels a huge distance in space without distorting its shape and without losing the information encoded in it. Contradictions in the notion of the photon and the fact that this notion is understood differently in different situations were observed time and again in the literature [1-3].

It is generally believed that convincing corroboration to the fact that light is made up of particles, i.e. the photons, is furnished by the photoelectric and Compton effects. At the same time, attempts were made to explain these phenomena on the base of undulatory concepts (see, e.g., [4-6]). Such attempts, however, suffer from an essential shortcoming: in order to explain these (and other) effects one is obliged to invent an individual mechanism for each effect, the mechanism being rather involved sometimes, whereas the explanation of all these effects with use made of the photonic concept alone indicates that a single mechanism lies at the root of the effects. Besides, the undulatory explanations mentioned sidestep often the fact that the photoelectric and Compton effects are observed practically without time delay [2,5].

The photon is one of the basic notions in quantum electrodynamics. In quantum electrodynamics, however, the photon is by no means a tiny particle. The photon is an elementary excitation of the electromagnetic field, the excitation that concerns the whole of the volume occupied by the system in question. Berestetskii et al. [7] underline that the wave function of a photon does not altogether signify the same thing as does the wave function of an ordinary particle in quantum mechanics, and the notion of coordinates of the photon has no physical meaning at all (see also [8], Sec. 5c). According to Bogoliubov and Shirkov [9], Sec. 6.2, by a particle in quantum field theory is meant an excitation of an oscillator, so that a $n^{th}$ excited state of the oscillator is referred to by convention as a state with $n$ particles, the photons in our case. From [9] it follows as well that, in a coordinate space, to the photon corresponds a plane wave $\exp[i(\omega t - \mathbf{k}\mathbf{x})]$ filling the overall volume of a big cube $V$. Nowhere in quantum electrodynamics does one resort to the concept of a point-like photon, and the photon in quantum electrodynamics is a wave pure and simple. In this connection Lamb [3] proposes to give up the use of the word 'photon' that often leads to confusion. An adequate terminology is always useful, of course, while a more important point emerges from the foregoing. Strange though it might appear at first thought, it is quantum electrodynamics that points to the fact that there must exist a wave explanation of the processes relevant to the photoelectric and Compton effects (quantum electrodynamics itself does not consider time development of the processes inasmuch as its methods are intended to calculate only the result of the processes).

The question of another sort arises in connection with the concept of the photon. It would seem rather strange that two very different physical realms such as an electromagnetic field and a material quantum system (say, an atom) are characterized by one and the same constant, namely, the Planck constant $\hbar$. Let us write down Bohr's relation relevant to emission or absorption of electromagnetic radiation by the quantum system:

$$\hbar\omega = E_n - E_m. \qquad (1.1)$$

On the left, one has the energy of a photon with angular frequency $\omega$. The energy levels $E_n$ and $E_m$ of the system on the right depend also upon the same constant $\hbar$. Besides, the surprising thing is that different quantum systems create and emit identical photons whatever the nature and structure of a quantum system may be. It is worth noting that the interaction between the electromagnetic field and the quantum system is characterized by another constant, the charge, that pertains to the quantum system alone. Moreover, spectral lines have a natural breadth various for different lines, so that the radiation contains a continuous set of frequencies in the vicinity of the frequency given by (1.1) rather than one frequency $\omega$.

The same quantum system can be in a nonstationary state, in which case its energy does not equal any $E_n$. No relation is known which could replace Bohr's rule of (1.1) when the system passes between nonstationary states or from a nonstationary state into a stationary one. At the same time, some radiation must exist in these situations too.

The preceding paragraph suggests that, before proceeding to the problem of the photon, it is necessary to revisit the nonstationary states in quantum mechanics insofar as the stationary states alone are usually discussed in detail. In this paper it will be shown that the resulting logical chain that invokes electromagnetic radiation and resonance phenomena leads to the conclusions that all effects ascribed to manifestation of the corpuscular properties of light can be completely explained on a base of the wave concept of light alone and therefore there are no photons in nature. The question as to why quantum electrodynamics employs the notion of a photon (and the notion is helpful there) will also be discussed in the concluding section.

## 2. Nonstationary states

Let us present first a brief summary of well-known facts that will be needed for what follows. A special role in quantum mechanics is ascribed to stationary states whose energy will be denoted as $E_n$ and the corresponding eigenfunctions as $\psi_n(q)$ where $q$ is a set of coordinates. We suppose that the system in question is isolated and the Hamiltonian does not depend on the time. For the sake of simplicity we assume $E_n$ to be discrete. If one knows $\psi_n(q)$, one can also write down the general solution of the Schrödinger equation for an arbitrary nonstationary state of the system (see, e.g., [10,11]):

$$\Psi(q,t) = \sum_n a_n e^{-\frac{i}{\hbar}E_n t} \psi_n(q), \qquad (2.1)$$

the coefficients $a_n$ satisfying the normalization condition

$$\sum_n |a_n|^2 = 1. \qquad (2.2)$$

The coefficients can be calculated if the initial wave function $\Psi(q,0)$ is known with use made of the fact that the functions $\psi_n(q)$ are orthonormal.

The mean energy $\overline{E}$ of the system in the nonstationary state described by the wave function $\Psi(q,t)$ is computed by one of the formulae

$$\overline{E} = \int \Psi^* \mathsf{H} \Psi \, dq = i\hbar \int \Psi^* \frac{\partial \Psi}{\partial t} dq, \qquad (2.3)$$

where $\mathsf{H}(p,q)$ is the Hamiltonian of the system and the Schrödinger equation has been used. Substituting (2.1) into the last expression of (2.3) yields

$$\overline{E} = \sum_n |a_n|^2 E_n. \qquad (2.4)$$

We can also calculate the mean square deviation from the mean energy:

$$\overline{\Delta E^2} \equiv \overline{(E - \overline{E})^2} = \int \Psi^* (\mathsf{H} - \overline{E})^2 \Psi \, dq = -\hbar^2 \int \Psi^* \frac{\partial^2 \Psi}{\partial t^2} dq - \overline{E}^2. \qquad (2.5)$$

In the case of (2.1) one has

$$\overline{\Delta E^2} = \sum_n |a_n|^2 E_n^2 - \overline{E}^2 . \tag{2.6}$$

If the system is in a stationary state where all $a_n = 0$ except for a coefficient $a_m = 1$, then (2.4) and (2.6) give $\overline{E} = E_m$, $\overline{\Delta E^2} = 0$. The energy of the system in this state has a strictly specified value. If the system is in a nonstationary state, its energy is not defined strictly although the mean energy is time independent. At the same time, (2.4) shows that the mean energy can have continuous values even in those cases in which the energy eigenvalues $E_n$ are discrete. This is due to the fact that the coefficients $a_n$ in (2.4) can be arbitrary, subject to the condition of (2.2) alone.

In order to see that these well-known facts entail unusual consequences it will be more informative to consider a case in which the infinite series of (2.1) can be represented in a finite form. As a characteristic example of the quantization of energy, we may take a harmonic oscillator. It is generally believed that the energy of the oscillator can assume only the values

$$E_n = \hbar \omega_c \left(n + \tfrac{1}{2}\right), \qquad n = 0, 1, 2, \ldots; \tag{2.7}$$

where $\omega_c = (k/m)^{1/2}$ is the classical angular frequency, $k$ is the force constant and $m$ is the mass. Let us consider, however, the one-dimensional oscillator proceeding on the time-dependent Schrödinger equation

$$i\hbar \frac{\partial \Psi}{\partial t} = -\frac{\hbar^2}{2m} \frac{\partial^2 \Psi}{\partial x^2} + \frac{1}{2} k x^2 \Psi . \tag{2.8}$$

In his textbook [11], Sec. 13, Schiff obtains a solution to this equation subject to the initial condition

$$\Psi(x,0) = \frac{\alpha^{1/2}}{\pi^{1/4}} e^{-\alpha^2 (x-a)^2 / 2}, \quad \alpha^2 = \frac{m\omega_c}{\hbar} . \tag{2.9}$$

If $a = 0$, this last function is the eigenfunction for the oscillator ground state ($n = 0$). In the case of (2.9), the series in (2.1) can be summed up. We shall not adduce the resulting Schiff's formula for $\Psi(x,t)$ [11] restricting ourselves by the probability density for the coordinate $x$:

$$\rho(x,t) \equiv |\Psi(x,t)|^2 = \frac{\alpha}{\sqrt{\pi}} e^{-\alpha^2 (x - a \cos \omega_c t)^2} . \tag{2.10}$$

According to this formula the wave packet oscillates without changing its shape about the point $x = 0$ with the amplitude $a$ and frequency $\omega_c$ just as a classical oscillator.

Let us calculate the mean energy of the oscillator in this state. Upon substituting Schiff's $\Psi(x,t)$ into the second formula of (2.3) with $q = x$, we obtain

$$\bar{E} = \frac{\hbar\omega_c}{2} + \frac{k a^2}{2}. \tag{2.11}$$

On account of (2.5), we may also compute the uncertainty for the energy:

$$\Delta E \equiv \sqrt{\overline{\Delta E^2}} = \sqrt{\hbar\omega_c}\sqrt{\frac{k a^2}{2}} = \hbar\omega_c\sqrt{\frac{k a^2}{2\hbar\omega_c}}. \tag{2.12}$$

It follows from (2.11) that the mean energy of the oscillator can take on arbitrary continuous values from $\hbar\omega_c/2$ and on. Curiously, the second term in (2.11) is of the same form as in the classical case. The sole dissimilarity from the classical case is in the presence of the zero-point energy $\hbar\omega_c/2$. If $ka^2/2 < \hbar\omega_c$, then $\Delta E < \hbar\omega_c$ (if $ka^2/2 \ll \hbar\omega_c$, then even $\Delta E \ll \hbar\omega_c$). In this case the energy of the oscillator that is not in the ground state is definitely less than the energy $E_1$ of the first excited state from (2.7).

Let us consider yet another example in brief. Let the normalized wave function of the oscillator at the initial instant be

$$\Psi(x,0) = \frac{\beta^{1/2}}{\pi^{1/4}} e^{-\beta^2 x^2/2} \tag{2.13}$$

with an arbitrary constant $\beta$. If $\beta \neq \alpha$ with $\alpha$ from (2.9) this function differs from the eigenfunction of the oscillator ground state only in width and height. Given (2.13) it is not difficult either to calculate the coefficients $a_n$ of (2.1) and sum up the series. The result is

$$\Psi(x,t) = \frac{\alpha}{\pi^{1/4}}\sqrt{\frac{2\beta}{A+B\zeta}}\exp\left(-\frac{i}{2}\omega_c t - \frac{\alpha^2 x^2}{2}\frac{A-B\zeta}{A+B\zeta}\right), \tag{2.14}$$

where $A = \alpha^2 + \beta^2$, $B = \alpha^2 - \beta^2$, $\zeta = e^{-2i\omega_c t}$. One can directly verify that this $\Psi(x,t)$ satisfies (2.8). The coordinate probability density is now of the form

$$\rho(x,t) = \frac{\alpha^2\beta}{\sqrt{\pi}}\sqrt{\frac{2}{\alpha^4+\beta^4+(\alpha^4-\beta^4)\cos 2\omega_c t}}\exp\left(-\frac{2\alpha^4\beta^2 x^2}{\alpha^4+\beta^4+(\alpha^4-\beta^4)\cos 2\omega_c t}\right). \tag{2.15}$$

Equation (2.15) shows that the centre of the wave packet stays put while the packet pulsates changing periodically its width and height. This movement of the quantum oscillator has no classical analogue. It is worthy of remark that such states are considered in studies on squeezed light [12].

Analogously to (2.11) and (2.12) we obtain now that

$$\bar{E} = \frac{\hbar\omega_c}{2}\left[1 + \frac{(\alpha^2-\beta^2)^2}{2\alpha^2\beta^2}\right], \quad \Delta E = \frac{\hbar\omega_c|\alpha^4-\beta^4|}{2\sqrt{2}\alpha^2\beta^2}. \tag{2.16}$$

Here again the energy can, depending on $\beta$, be arbitrary beginning with $\hbar\omega_c/2$.

It should be observed that the average value of the coordinate of the quantum harmonic oscillator in any state is always given by the formula [13]

$$\bar{x} = a\cos(\omega_c t + \varphi), \qquad (2.17)$$

analogous with the classical formula. In particular, the solution of (2.10) corresponds completely with (2.17). As to (2.15) one has $a = 0$. For the stationary states, $a = 0$ as well.

Before proceeding to further examples, some preliminary comment upon the results obtained is in order at this point. From the above examples it follows that the quantum oscillator is capable of absorbing or emitting energy in continuous and arbitrary amounts if its nonstationary states are taken into account. In order to have the values of (2.7) it is necessary to 'prepare' such an initial wave function $\Psi(x,0)$ that is equal to $\psi_n(x)$ with an absolute precision (for example, one must obtain exactly $a = 0$ in (2.9) or $\beta = \alpha$ in (2.13)), which is next to impossible. Practically any external action will result in various $a_n$'s in (2.1) that will be different from zero, and the state will not be stationary. The examples of (2.10) and (2.15) are only two of an infinite number of possibilities for the nonstationary states of the quantum oscillator. From this point of view, not only do the stationary states of a quantum system fail to stand out from all possible states but also the probability of their observation in nature or in experiment should be factually nil.

Another strange fact follows from the above analysis. According to (2.10), (2.15) and (2.17) the quantum oscillator will oscillate infinitely long without showing any tendency to pass into a stationary state, in particular, into the ground state. The same is evident from (2.1) in the general case because all terms in the sum regarded as functions of the time $t$ oscillate with constant amplitudes. At the same time, from experiment and quantum electrodynamics it is well known that even the lifetime of stationary (excited) states is finite. In this connection, the fact that the life time of the nonstationary states should be infinite looks absurd. The problem may be resolved as follows. The wave function of (2.1) is a superposition of stationary-state wave functions. Once each of the excited stationary states has a finite life time, the life time of the superposition must be finite, too. This last life time should be of the order of the life time of the most long-lived excited stationary states. However, if this is the case, the stationary states will not be outstanding as compared with the nonstationary ones in this respect either.

We turn now to two examples relevant to a hydrogen atom. This is rather a difficult matter to sum up the series of (2.1) in this case if an initial function $\Psi(q,0)$ is given. For this reason we proceed in another way. In (2.1), we take only two coefficients, say, $a_1$ and $a_2$, different from zero and set all other coefficients equal to zero. We shall lean upon formulae of [10] for a hydrogen atom implying atomic units defined in [10]. To begin with, for $\psi_1(\mathbf{r})$ we take the eigenfunction relevant to the quantum numbers $n = 1$ and $l = m = 0$, whereas for $\psi_2(\mathbf{r})$ the eigenfunction relevant to $n = 2$ and $l = m = 0$. Then (2.1) yields

$$\Psi(r,t) = \frac{1}{\sqrt{\pi}} e^{\frac{1}{2}it} \left[ a_1 e^{-r} + \frac{a_2}{2\sqrt{2}} \left(1 - \frac{r}{2}\right) e^{-\frac{1}{2}r - \frac{3}{8}it} \right], \tag{2.18}$$

in which the coefficients $a_1$ and $a_2$ supposed to be real for simplicity's sake are related by $a_1^2 + a_2^2 = 1$ in view of (2.2). If need be, Eq. (2.18) permits one to find $\Psi(r,0)$.

The probability density of the coordinate $r$ is

$$\rho(r,t) \equiv |\Psi(r,t)|^2 = \frac{1}{\pi} e^{-r} \left[ a_1^2 e^{-r} + \frac{a_2^2}{8}\left(1 - \frac{r}{2}\right)^2 + \frac{a_1 a_2}{\sqrt{2}}\left(1 - \frac{r}{2}\right) e^{-\frac{1}{2}r} \cos\tfrac{3}{8}t \right], \tag{2.19}$$

from which it is seen that the atom pulsates remaining spherically symmetric. The mean energy of the atom can be found directly from (2.4):

$$\bar{E} = -\frac{1}{2} + \frac{3}{8} a_2^2, \tag{2.20}$$

where the coefficient $a_1$ is excluded and $0 \le a_2 \le 1$. One sees that, in the case of the atom, the mean energy varies continuously as well, from $E_1 = -1/2$ to $E_2 = -1/8$ in this example.

Next we consider another example by taking, for $\psi_1(\mathbf{r})$, the eigenfunction relevant to the quantum numbers $n = 1$ and $l = m = 0$, and for $\psi_2(\mathbf{r})$ the eigenfunction relevant to $n = 2$ and $l = m = 1$. In this instance Eq. (2.1) gives

$$\Psi(\mathbf{r},t) = \frac{1}{\sqrt{\pi}} e^{\frac{1}{2}it} \left[ a_1 e^{-r} + \frac{ia_2}{8} r \sin\theta \, e^{-\frac{1}{2}r - \frac{3}{8}it + i\varphi} \right]. \tag{2.21}$$

The probability density of coordinates is now

$$\rho(\mathbf{r},t) = \frac{1}{\pi} e^{-r} \left[ a_1^2 e^{-r} + \frac{a_2^2}{64} r^2 \sin^2\theta + \frac{a_1 a_2}{4} r \sin\theta \, e^{-\frac{1}{2}r} \sin\left(\tfrac{3}{8}t - \varphi\right) \right]. \tag{2.22}$$

The maximum of the density falls on the angle $\varphi = 3t/8 - \pi/2$ if one assumes that $a_1 a_2 > 0$ for the sake of definiteness. Therefore the atom is not axially symmetric (the hydrogen atom stationary states are axially symmetric), and the probability distribution rotates around the $z$-axis with an angular velocity of $d\varphi/dt = 3/8$. The energy here is given by (2.20) as before.

Of some interest is the angular momentum $\mathbf{l}$ of the atom in this example. The mean projection of the angular momentum on the $z$-axis and the mean $\mathbf{l}^2$ expressed in terms of $\hbar$ are given by [10]

$$\bar{l}_z = -i \int \Psi^* \frac{\partial \Psi}{\partial \varphi} d\mathbf{r}, \quad \overline{\mathbf{l}^2} = -\int \Psi^* \left[ \frac{1}{\sin^2\theta} \frac{\partial^2 \Psi}{\partial \varphi^2} + \frac{1}{\sin\theta} \frac{\partial}{\partial \theta}\left(\sin\theta \frac{\partial \Psi}{\partial \theta}\right) \right] d\mathbf{r}. \tag{2.23}$$

Upon substituting (2.21) we obtain

$$\bar{l}_z = a_2^2, \quad \overline{\mathbf{l}^2} = 2a_2^2. \tag{2.24}$$

The quantities $\bar{l}_z$ and $\overline{\mathbf{l}^2}$, like the energy, take on continuous values with $0 \leq |a_2| \leq 1$. It may be added that $\bar{l}_x = \bar{l}_y = 0$ in this case.

Thus we see that, analogously with the quantum oscillator, the energy of the atom is not quantized, being continuous. By taking other values of *n*, one could obtain values of energy differing from the ones given by (2.20).

We end this section by a remark. If, in (2.1), one takes different non-zero $a_i$'s that are relevant, however, to one and the same energy level $E_n$ (in case the level is degenerate), then the common factor $\exp(-iE_n t/\hbar)$ will factor out from the sum. The quantity $|\Psi(q,t)|^2$ in this case will be time-independent just as in a stationary state. Such a state, however, does not correspond to the standard definition of the stationary state because it will contain several eigenfunctions $\psi_i(q)$. At the same time, the quantity $E_n$ factors out from the sum in (2.4) as well, and we shall have $\bar{E} = E_n$ owing to (2.2). In a like manner, Eq. (2.6) yields $\overline{\Delta E^2} = 0$, which amounts to saying that the energy has a strictly specified value in these peculiar states that are not eigenstates.

In this context, let us revert to the hydrogen atom as example. At a given principal quantum number $n \neq 1$, there are eigenfunctions with different orbital quantum numbers *l* and magnetic quantum numbers *m*. If we place all these eigenfunctions with different *l* and *m* in (2.1) (the index *n* should be replaced by *nlm*), we shall obtain a state of the hydrogen atom in which the energy $E_n$ will be strictly fixed together with *n* and different probabilities will be time-independent. The angular momentum, however, will be quantized neither in magnitude nor in space, and can take on continuous values depending on the coefficients $a_{nlm}$ analogously with (2.24).

## 3. Electromagnetic radiation

The preceding section shows that, according to quantum mechanics taken alone, the stationary states do in no way stand out as compared to the nonstationary ones, and moreover the stationary states are difficult if not impossible to realize. At the same time, a wealth of experiments accumulated in physics through the 20[th] century point to a peculiar role of the stationary states and to the fact that the stationary states alone manifest themselves in actual practice almost always.

To resolve this paradox, account must be taken of the fact that all usual particles create electric and magnetic fields. For example, a neutron that has no charge possesses a magnetic moment and thus produces a magnetic field; neutral atoms are made up of charged particles.

After this remark we revert to the nonstationary states reasoning first in classical terms and considering systems of charges particles for brevity sake. Any nonstationary process implies that the particles of a system have some accelerations or other because movement of all the particles with a constant common velocity means a stationary process (interacting particles cannot move with different and constant velocities). Every accelerated charge always emits electromagnetic waves, however. Therefore, any nonstationary process in a system containing charged particles is accompanied with electromagnetic radiation. A generalization of this statement valid also for quantum systems follows from the Maxwell equations according to which a time-variant source of electric or magnetic field generates electromagnetic waves. Note that the Maxwell equations lie in the heart of quantum electrodynamics. In the case of a quantum system, the density of charge or of magnetic moment is proportional to $|\Psi(q,t)|^2$. When the system is in a nonstationary state, the quantity $|\Psi(q,t)|^2$ depends upon the time and the system produces electromagnetic radiation that carries away energy. In this regard the stationary states possess a unique property, namely, if the system is in a stationary state, the electromagnetic radiation is lacking for $|\Psi(q,t)|^2$ is time-independent. As a result, the quantum system radiating the energy will inevitably pass from any nonstationary state into one of the stationary.

We can estimate the time it takes for the transition. This time should be in magnitude of the order of a characteristic time present in the time-dependent Schrödinger equation. If a hydrogen atom is taken as example, the natural units are Hartree's units employed in the preceding section. In this atomic system of units, the unit of time is $\tau_q = \hbar^3/me^4 = 2.4 \cdot 10^{-17}$ s [10]. As long as electrodynamical processes attended by the electromagnetic radiation come into play too, one ought to estimate the time necessary for development of these processes. This last time $\tau_{ed}$ is of the order of the time it takes for light to travel the region where the processes evolve. In the case under discussion, the size of the region will be something like the diameter of the hydrogen atom, that is, $\approx 1$ Å, which gives $\tau_{ed} \approx 3 \cdot 10^{-19}$ s. In the general case, $\tau_{ed} \sim \alpha \tau_q$ where $\alpha \approx 1/137$ is the fine structure constant, which is obvious from dimension considerations since $\tau_{ed}$ should contain the velocity of light $c$ in the denominator while there is only one suitable dimensionless constant, namely, $\alpha = e^2/(\hbar c)$. It is worthwhile also to estimate the time relevant to the emission of radiation by accelerated charges, which can be readily done with the help of classical electrodynamics. The equation of motion of an electron with account taken of the radiation reaction is

$$m\frac{d\mathbf{v}}{dt} = \mathbf{F} + \frac{2e^2}{3c^3} \cdot \frac{d^2\mathbf{v}}{dt^2}, \tag{3.1}$$

in which **F** is an external force acting on the electron and the last term represents the radiation reaction [14]. Calculation with use made of quantum electrodynamics gives a similar result [15]. Comparing the dimensions of the first and last entries in (3.1) shows that this problem contains a characteristic time $\tau_{rd} = e^2/(mc^3)$ whence $\tau_{rd} \sim \alpha^2 \tau_{ed}$. As long as $\tau_{rd} \ll \tau_{ed}$, the processes leading to the radiation emission by accelerated electrons are ahead of other electrodynamical processes. At the same time $\tau_{ed} \ll \tau_q$, so that no electrodynamical processes retard the quantum ones and the atom passes from a nonstationary state to a stationary during the time of order $10^{-17}$ s. As was mentioned in the preceding section, if the stationary state is excited, its life time is finite; however this last life time that is of the order $10^{-8}$ s for optical transitions is many orders of magnitude greater than $\tau_q$.

At the end of the preceding section, the case was considered of states in which $|\Psi(q,t)|^2$ is time-independent but which are not energy eigenstates. These states may occur only if the relevant energy level is degenerate. A strict degeneracy is possible, however, for a fully isolated atom alone because external fields, as a rule, lift the degeneracy. For there can be no truly isolated atoms in nature, in fact there are no states in which $|\Psi(q,t)|^2$ is time-independent except for the energy eigenstates.

As a result, we see that in a matter of a very short lapse of time the quantum system passes from any nonstationary state into a stationary one. This is the reason why the stationary states alone are factually observed in actual practice, although an external action generally transfers the system first to a nonstationary state. By the way, the foregoing permits one to clarify the question as to why a classical apparatus that performs energy measurements gives only eigenvalues $E_n$, the fact that is postulated in quantum mechanics but is not explained anyhow. No classical device is capable to respond to a process that lasts for $\sim 10^{-17}$ s. As a consequence, when the device begins to respond, the quantum system will be already in a stationary state, and the device will generate characteristics of this last state. It is logical to suppose that the transition is most likely to occur to eigenstates for which the coefficients $a_n$ in (2.1) are greatest in modulus because the change in the energy will be minimal in this case according to (2.4). This being so, the apparatus gives the value $E_n$ with a probability of $|a_n|^2$, which is postulated in quantum mechanics.

There may be a feeling of dissatisfaction due to the fact that in order to explain principles of quantum mechanics one is obliged to resort to another branch of physics, electrodynamics. It is worth recalling, however, that it is just electrodynamics, more precisely the existence of electromagnetic radiation, that points to inadequacy of classical mechanics for description of atomic phenomena (emitting the radiation the rotating electron should fall on the nucleus if it obeyed the laws of classical mechanics), although this does nowise follow from classical mechanics itself. If electrodynamics is so crucial as to classical mechanics when applied to atomic phenomena, it would appear natural that electrodynamics affects quantum mechanical processes as well. A distinguishing feature of quantum mechanics is that it allows of the existence of stationary states that have no analogue in classical mechanics except for the state of static equilibrium that is impossible, by the way, in case electric forces alone act in the system according to Earnshaw's theorem [16]. What is more, the following well-known evidence may be cited as to the manifestation of laws of electrodynamics in quantum phenomena. Only (quantum) electrodynamics enables one to find out that the life time of excited stationary states is finite. This does not at all follow from quantum mechanics itself. Once electromagnetic effects tell essentially upon properties of the stationary states, their role for nonstationary states must be no less crucial.

The aforesaid may be expressed in other words. By and large, the finite life time of the stationary states is due to the fact that the simultaneous set of the Dirac and Maxwell equations employed by quantum electrodynamics has no stationary solutions (except for the ground state). The Dirac equation by itself does admit stationary solutions. The instability of these solutions is induced by the action of the electromagnetic field created by the electron upon the electron itself (the self-field leads to a factor $\exp[-\Gamma_n t/(2\hbar)]$ with $\Gamma_n > 0$ by which the eigenfunctions of the excited states should be multiplied, so that the states will not be strictly stationary). This effect is rather subtle, which results in a sufficiently long life time of the stationary states. In the case of the nonstationary states, from the Schrödinger equation (or from the Dirac equation in a relativistic problem) it follows directly that the density of charge depends upon the time. This being so, the Maxwell equations show immediately that there must be electromagnetic radiation. This effect is well pronounced, and the life time of the nonstationary states is very short. Thus, the properties of the stationary states as well as those of the nonstationary ones cannot be understood without invoking electrodynamics.

The foregoing shows that we revert to the usual understanding of a special role played by the stationary states, i.e., the energy eigenstates. At the same time the above consideration leads inevitably to an unusual conclusion as well, namely, the electromagnetic radiation relevant to the transition from a nonstationary state to a stationary cannot by emitted by quanta, it must be emitted continuously. This is due to the fact that according to Sec. 2 the energy of a nonstationary quantum system can be arbitrary, whereas the energy quantum is fixed and equal to $\hbar\omega$. For example, the last term in (2.11) depends solely on the amplitude $a$ determined by initial conditions and is not connected with any frequency just as in the classical case. If one wants to conserve the idea of photons, one should admit rather a strange thing that there are two types of electromagnetic radiation: a continuous radiation and a particle-like radiation relevant to (1.1). This is an added reason, besides the ones mentioned in Introduction, for revising the concept of photons.

One further question arises in the context of the above deductions. In this section we discussed what happens with a quantum system after its receiving an amount of energy. This discussion does not touch upon another consequence that follows from the results of the preceding section, namely, a quantum system even with discrete energy eigenvalues is able to absorb energy in arbitrary amounts upon passing first to a nonstationary state whose energy can vary continuously. At the same time there exists, for example, the famous Franck-Hertz experiment [17] from which it follows a seemingly unambiguous conclusion that the quantum system (an atom in that experiment) is capable of absorbing energy only in definite discrete packets in accord with (1.1). This question will be taken up in the next section.

## 4. Resonance interaction

The remark at the end of the preceding section suggests that the time is right to analyse the question as to what happens when a quantum system receives energy. We shall employ perturbation theory upon presuming that the interaction of the system with a source of energy is sufficiently weak. For the sake of simplicity we assume that the life time of occurring stationary states is infinite.

First we consider the case in which our system is acted upon by a time-periodic perturbation of frequency $\omega$. Let the system be initially in a stationary state where $a_m = 1$ and $a_n = 0$ with $n \neq m$ in (2.1). With use made of Landau and Lifshitz's [10], Sec. 40, formulae in a slightly different notation we obtain that, as a result of the action of the perturbation, the coefficients $a_n$ become, in a first approximation,

$$a_n^{(1)} = -\frac{F_{nm}}{\hbar(\omega_{nm} - \omega)} e^{i(\omega_{nm} - \omega)t} - \frac{F_{mn}^*}{\hbar(\omega_{nm} + \omega)} e^{i(\omega_{nm} + \omega)t}, \qquad (4.1)$$

where $n \neq m$ and $\omega_{nm} = (E_n - E_m)/\hbar$. We shall not write down Landau and Lifshitz's formula for $F_{nm}$ because our prime interest here is with the qualitative aspect of the problem. It is important for us only that the quantity $F_{nm}$ is proportional to the amplitude of the perturbation.

It follows from (4.1) that, in case the denominators are not close to zero, the coefficients $a_n$ are small by virtue of the weakness of the perturbation. The system passes into a nonstationary state since different $a_n$'s in (2.1) are nonzero. Only $a_m \approx 1$ while the energy of the system will differ little from $E_m$ in view of (2.4). When the action of the perturbation is terminated, the system will return to the initial stationary state within a very short lapse of time ($\sim 10^{-17}$ s) owing to the electromagnetic radiation considered in the preceding section.

The situation changes drastically if one of the denominators in (4.1) is close or equal to zero, in which case (4.1) becomes invalid. This case is considered by Landau and Lifshitz [10] in the problem relative to their section 40. We shall suppose that there is an exact resonance, so that $\omega = \omega_{nm}$ or $\hbar\omega = E_n - E_m$ for some $n$. Then from that problem it follows that

$$|a_n|^2 = \tfrac{1}{2}[1 - \cos(2|\eta|t)], \qquad (4.2)$$

where $\eta = F_{nm}/\hbar$. Equations (4.1) and (4.2) differ essentially in character. In contradistinction to (4.1), the coefficient $a_n$ according to (4.2) is no longer proportional to the amplitude of the perturbation and its modulus can take on any values between 0 and 1 depending on the time $t$. The periodic time-dependence of $|a_n|$ as given by (4.2) may be readily understood: when the probability of the system's being in the upper energy level $E_n$ approaches unity, the perturbation induces the system to descend onto the lower level $E_m$ and so on.

Let now there be an array of atoms in an eigenstate with the energy $E_m$ acted upon by a periodic perturbation of frequency $\omega$. If there is no resonance, after cessation of the perturbation action all the atoms will return to the initial state in a time of ~$10^{-17}$ s (see above). In case the resonance occurs, when $\hbar\omega = E_n - E_m$, after cessation of the perturbation action some atoms will remain in the $m^{\text{th}}$ eigenstate, the probability of which being equal to $|a_m|^2 = 1 - |a_n|^2$; while the others will find themselves in the $n^{\text{th}}$ eigenstate with a probability of $|a_n|^2$ given by (4.2) with some $t$. All of this looks as though the atoms were able to absorb energy only in packets of $E_n - E_m$ while, on the other hand, the packets were equal to $\hbar\omega$.

It should be emphasized that, when deriving (4.1) and (4.2), the perturbation is implied to be continuous and not consisting of any discrete portions (quanta). The perturbation is not even considered to mean an electromagnetic wave. The emerging relation $\hbar\omega = E_n - E_m$ is akin to Bohr's relation of (1.1). In the present case, however, the constant $\hbar$ in $\hbar\omega$ appears owing to properties of the quantum system alone and is not due to any electromagnetic radiation. Bohr's relation would be more properly recast as

$$\omega = \frac{E_n - E_m}{\hbar}, \qquad (4.3)$$

thus defining the resonance frequency and nothing more.

In this connection it is instructive to revert to the example of the harmonic oscillator considered in Sec. 2. The minimal value of the difference $E_n - E_m$ for the oscillator is $\hbar\omega_c$ with reference to (2.7). Placing this value in (4.3) we see that the resonance occurs when $\omega = \omega_c$, which is completely analogous with the classical case. Here, however, the analogy ends up. If there is no dissipation of energy, the amplitude of a classical oscillator tends to infinity when $\omega \to \omega_c$, whereas in the quantum case the resonant oscillator passes periodically from the $m^{th}$ eigenstate into the $n^{th}$ eigenstate and vice versa in view of (4.2). It should be added that in the general case the quantum resonance occurs if $\omega = k\omega_c$ with $k = 1, 2, 3,\ldots$.

Another important feature of the resonance phenomena can be found out by analyzing the scattering of light by bound electrons, for example, by the electrons in an atom. This question is considered in [18], Sec. 4.2. It should be remarked that, although Akhiezer and Berestetskii [18] speak about the scattering of photons, they in no way use the fact that the photons are particles (see Introduction in the present paper). The total scattering cross section in question is given by Eq. (4.2.45) of [18] [in the edition of 1965 this is Eq. (35.15)], namely,

$$\sigma = \frac{2\pi}{\omega^2} \frac{\Gamma_n^2}{(E_n - E_m - \omega)^2 + \Gamma_n^2/4}, \quad (4.4)$$

where $\Gamma_n$ is the resonance width. Note that this equation is written using units where $\hbar = c = 1$. We have put $j_s = j_1$ for simplicity sake, and made the replacement $1 \to m$ and $s \to n$ in accord with the foregoing. If the quantity $E_n - E_m - \omega$ is not close to zero, the cross section $\sigma$ is rather small being proportional to $\Gamma_n^2/\omega^2$ whereas $\Gamma_n \ll \omega$. If, however, $E_n - E_m - \omega = 0$ [cf. (4.3)], that is to say, if the resonance occurs, we shall have

$$\sigma = \frac{8\pi}{\omega^2} = \frac{2}{\pi}\lambda^2, \quad (4.5)$$

where $\lambda = 2\pi/\omega$ is the wavelength of the incident wave in the units used. For visible light, $\sigma \sim (5\cdot 10^{-5} \text{ cm})^2 \sim 10^{-9} \text{ cm}^2$ while the visual cross section of the atom is of the order $(10^{-8} \text{ cm})^2 = 10^{-16} \text{ cm}^2$ being extremely small as compared with $\sigma$. In order to act on the electrons in the atom light must penetrate into the atom. At the same time, not only a small fraction but all this light flux described by $\sigma$ participates in the processes that take place inside the atom. Therefore, when the resonance occurs, the atom deflects the electromagnetic wave and gathers it into a narrow beam that penetrates the atom, the deflection being extended over enormous distances.

Thus we can see that a resonator is able to suck in the incident wave from a tremendous region of space, redistributing energy flows and so accumulating energy. This can be attributed to the presence of positive feedback in the system made up of the resonator and the incident wave. Since the wave contracts into a narrow beam, the amplitude of the wave should grow considerably in the neighbourhood of the resonator.

When deducing (4.2), it is assumed that the resonator has no effect on the perturbation as long as the perturbation is fixed. However, under these conditions too, the capability of the resonator for accumulating energy manifests itself. The time it takes for the resonator to absorb an energy packet of $E_n - E_m$ when $|a_n| = 1$ in (4.2) is $t = \pi/(2|\eta|)$. The weakness of the perturbation (the smallness of $|\eta|$) affects only the time necessary to accumulate the energy. When the resonator can redistribute energy flows by feedback as is the real situation with the atoms, the time taken to accumulate the energy should decrease appreciably

Having considered the periodic perturbation we now turn to the case in which the quantum system interacts with an impacting particle. The above results can be applied directly to this case too if account is taken of the undulatory properties of the particle. Let the particle be described by a plane de Broglie wave, in which case the frequency $\omega$ and the energy of the particle $E$ are related by $\omega = E/\hbar$ [10,11]. Upon substituting this $\omega$ into the resonance condition of (4.3), we see that the energy can be transmitted efficaciously from the particle to the system only if $E = E_n - E_m$.

Now it is not difficult to explain the Franck and Hertz experiment. Contrary to the usual assertion, the inelastic collisions of electrons with atoms, in which a part of the energy of an electron passes into the internal energy of a nonstationary atom, can occur when the energy of the electrons is arbitrary, be it even very low. In the latter case, however, the energy transfer is ineffective and the electrons practically do not lose their velocity after the collisions. Only when the energy of the electrons increases and becomes equal to $E_1 - E_0$ where $E_0$ is the energy of the ground state of the atom and $E_1$ of the first excited state, do the electrons begin to transmit their energy to the atoms efficaciously (see above). As a consequence the atoms excite themselves and commence to emit radiation while the energy of the electrons drops markedly. This is just the essence of the results obtained by Franck and Hertz (the current-voltage curve in the Franck-Hertz type experiments is a typical resonance curve with several resonances).

Resonance interactions are familiar in quantum physics. Let us cite an example that is relevant to the situation in which a colliding particle and a target can form a bound system that decays later on. In nuclear physics, such processes are known as nuclear reactions that proceed with formation of an intermediate compound nucleus. These processes are described by the Breit-Wigner formulae [10,13]. If the energy of the incident particle is close to one of the energy levels of the compound system, that is, if there occurs resonance, then the reaction cross section rises sharply. Blokhintsev [13], Sec. 81, adduces an example of $_{54}Xe^{135}$ nuclei where the resonance cross section of neutron scattering is $10^5$ times greater than the area of the geometrical cross section of the target nucleus. For the compound nucleus to form, the neutron must penetrate into the $_{54}Xe^{135}$ nucleus because nuclear forces are short-range and act only within the nucleus. At the same time, when the resonance happens, the enormous enhancement of the reaction cross section indicates that the neutrons, which should for the most part pass past the nuclei at distances from them far exceeding their size, do form compound nuclei. This seeming contradiction can be resolved if account is taken of wave properties of the neutrons, which is supported by the fact that the resonance cross section depends upon the neutron wave length analogously to (4.5) [13]. The target nucleus redirects the neutronic wave gathering it into a narrow beam that penetrates the nucleus. Here again the resonator is capable to suck in the incident wave from a tremendous region of space.

The above properties of resonators are familiar in acoustics. The amplitude of vibration in the hole of an acoustic resonator is many times greater than the amplitude of the acoustic field the resonator is placed in [19], which suggests essential redistribution of the incident acoustic wave and sucking the wave by the resonator because the latter amplifies sound using the energy of the incident wave alone. It is interesting that the wavelength relevant to low-frequency modes of Helmholtz's resonator far exceeds the size of the resonator. Analogously, atoms are essentially smaller in size than the wavelength of a light wave with which they can come into resonance.

Yet another example of the peculiarities of the resonance phenomena is furnished by lasers. The energy level population inversion is not sufficient for efficacious operation of a laser. The laser must contain an optical resonator. Only if there occurs resonance in the resonator, do energy flows in the laser redistribute themselves essentially so that the energy is efficaciously pumped from the active laser medium and concentrated in the laser beam. Here again we meet with the phenomenon of redistribution and concentration of energy flows owing to the resonance, which results in a considerable amplification of the wave intensity.

Equation (4.1) can be applied also to radiation of energy by a quantum system. The initial energy $E_m$ in this instance is higher than the final energy $E_n$, so that it is the second denominator in (4.1) that vanishes if $\hbar\omega = E_m - E_n$. Physically, the process of radiation may be conceived as follows. The system cannot remain in an excited state for long. The radiation the system begins to emit tending to diminish its energy acts on the system itself. Most efficiently emitted will be the radiation with a frequency $\omega$ at which the second denominator in (4.1) vanishes, that is to say, at which $\hbar\omega = E_m - E_n$. We see that the resonance phenomena manifest themselves not only when a quantum system absorbs energy but also when it emits energy. It is worth adding that the idea of resonance in quantum processes was exploited by Schrödinger [20] when discussing quantum jumps.

Closing this section we remark that the results of the present and preceding sections point to a dual role played by the stationary states. On the one hand, within a very short lapse of time the quantum system passes from a state with arbitrary energy into a stationary state. On the other hand, energy is most efficaciously received or released by the quantum system in packets equal to $E_n - E_m$. Besides, if the energy is absorbed or emitted in the form of a wave, the efficient frequency of the wave will be $\omega = |E_n - E_m|/\hbar$.

## 5. The photoelectric and Compton effects

In the preceding section we saw that Bohr's relation of (1.1) is merely the condition of a resonance and has nothing to do with the concept of a photon. In this section we shall show that the same resonance phenomena can explain the photoelectric and Compton effects and some other effects and moreover all peculiarities of these effects that are usually ascribed to particle-like photons.

A wave explanation of the photoelectric effect follows immediately from the results of the preceding section. The photoelectric effect signifies that one of the electrons of a metal comes into resonance with the incident light wave. The resonant electron efficaciously sucks in the energy of the wave until, so to speak, the resonator breaks down, that is to say, until the electron escapes from the metal. In this situation, the initial state of the electron pertains to a discrete spectrum while the final one to a continuous spectrum. In the resonance condition $\hbar\omega = E_n - E_m$, the energy difference $E_n - E_m$ is now equal to the work function $W$ plus the kinetic energy of the electron outside the metal, so that

$$\hbar\omega = W + \frac{mv^2}{2}. \tag{5.1}$$

We have obtained Einstein's photoelectric equation that is usually deduced on the presumption of the particle nature of light, whereas we did not resort to the concept of a photon.

In attempting to understand the photoelectric effect on undulatory grounds, the most incomprehensible point is as to how even a very feeble electromagnetic wave is able to eject an electron from a metal and impart an appreciable velocity to the electron. The examples of the preceding section (resonance scattering, acoustic resonators, lasers) suggests that the original incident wave intensity has nothing to do with the actual wave intensity in the vicinity of the resonant electron, the latter intensity being substantially higher than the former. When sucking in the wave, the resonant electron sculptures its own electromagnetic wave field that is sufficiently strong to eject the electron from the metal. No wave amplitude can enter into (5.1) because Eq. (5.1) is a resonance condition that must not contain any amplitude.

We turn now to the second seemingly incomprehensible point, the point concerning the negligible time delay in the course of the photoelectric effect (see Introduction). All processes in an electromagnetic wave run with the velocity of light (cf. Sec. 3). Therefore, distorting the incident wave and sucking in the wave by the resonant electron takes place with the light velocity as well, the more so as the own 'quantum mechanical' time of the resonator is extremely small ($\tau_q \sim 10^{-17}$ s for a hydrogen atom according to Sec. 3). As a result, the photoelectric effect happens in a lapse of time that must be shorter than the time it takes for the light to travel the distance from the source of light to the resonant electron. In other words, photoelectrons may appear almost immediately the light is switched on, which is observed in experiment.

Calculation of the photoelectric effect can be carried out with use made of the perturbation theory for transitions from discrete-spectrum states to continuous-spectrum ones in response to a periodic action (see, e.g., Sec. 42 of [10]). In that section, one can find a formula for the transition probability per unit time in the present case ($E_n = E$, $E_m = E_i$):

$$w_{Ei} = \frac{2\pi}{\hbar} |F_{Ei}|^2. \tag{5.2}$$

The quantity $F_{Ei}$ is proportional to the amplitude of the perturbation (cf. (4.1)), which amounts to saying that $|F_{Ei}|^2$ is proportional to the intensity of light. Hence from (5.2) it follows one of the laws of the photoelectric effect according to which the number of photoelectrons ejected from a metal per unit time is proportional to the intensity of incident light. We see that the light intensity enters only into (5.2) and not into (5.1).

A comprehensive calculation of the photoelectric effect in complete analogy with the above argumentation can be found, e.g., in [13], Sec. 95, (see also [11] where a similar approach is referred to as the semiclassical treatment of radiation). Although Blokhintsev [13] holds to the idea of photons, when carrying out the calculation he does nowise resort to this idea and represents the vector potential of the light wave in the usual form

$$\mathbf{A} = \tfrac{1}{2}\mathbf{A}_0\, e^{i(\omega t - \mathbf{k}\mathbf{r})} + \tfrac{1}{2}\mathbf{A}_0\, e^{-i(\omega t - \mathbf{k}\mathbf{r})}. \tag{5.3}$$

He obtains a relation of the type (5.1) as a condition of the resonance. On the basis of an equation analogous with (5.2), he calculates also the angular distribution of the photoelectrons that is in accord with experiment. In quantum electrodynamics, calculation of the photoelectric effect is carried out along similar lines [7], Sec. 56, and again no properties characteristic of a photon are utilized.

Two remarks need to be made here. In the cited calculations of [13] and [7], the above mentioned distortion of the wave by the resonant electron is not taken into account as long as the vector potential is considered to be fixed according to (5.3). When calculating, however, the probability of a transition with use made of quantum mechanical rules, one is to take only the initial and final states of the system. For this reason, in a formula of the type (5.2), parameters of the light wave at infinity are implied rather than the parameters of the electromagnetic field near the resonator. As a case in point we may cite also the theory of particle scattering in which one analyses only the asymptotic form of the incident and scattered waves relevant to the particles at infinity [10,11,13], whereas the exact form of the wave function near the scattering centre is not required.

We are coming now to the second remark. In actual sources of light employed in experiment, the light is emitted by separate atoms. An atom emits light in the form of a wave train whose dimensions are finite inasmuch as the energy radiated is finite and equal to $E_m - E_n$. The wave train can move in a particular direction as is the case with stimulated emission. Apart from the radiation of this sort, there exists radiation relevant to transitions from nonstationary states to stationary ones as discussed in Sec. 3. The last type of radiation should be made up of short pulses since the relevant emission time is brief (~$10^{-17}$ s for a hydrogen atom). For these reasons there are no ideal plane waves with constant amplitude in nature. Any real electromagnetic wave is nonuniform to some degree or other, the nonuniformity distribution being time-dependent. These nonuniformities play a minor part if the intensity of light is sufficiently high, whereas in the case of light waves of weak intensity used in some experiments the nonuniformities may manifest themselves markedly. For example, the photoelectric effect is most likely to occur at a point of the metal where the amplitude of the incident light wave is a maximum at a given instant.

The same resonance interactions can also explain the Compton effect, that is, scattering of electromagnetic radiation off free electrons. As distinct from the photoelectric effect, both the initial and final states of the electron pertain to the continuous spectrum. Besides, an important role in the Compton effect is played by the law of conservation of momentum (in the photoelectric effect a portion of the momentum is transferred to the metal).

In this connection there is a need to discuss the momentum of an electromagnetic wave. When considering a plane electromagnetic wave, Landau and Lifshitz [14] observe that the energy $W$ of the wave and its momentum $p$ are interrelated by $W = cp$. Consequently, to an energy packet of $\hbar\omega$ corresponds the momentum $p = \hbar\omega/c = \hbar k$ where the wavevector $\mathbf{k}$ is introduced ($|\mathbf{k}| \equiv k = \omega/c$). This relation can be rewritten in a vectorial form, $\mathbf{p} = \hbar\mathbf{k}$, as long as the two vectors $\mathbf{p}$ and $\mathbf{k}$ point to the same direction. The noteworthy fact is that the relation $\mathbf{p} = \hbar\mathbf{k}$ is akin to the formula for the momentum of a photon although we utilized only classical electrodynamics of [14]. The constant $\hbar$ appears in $\mathbf{p} = \hbar\mathbf{k}$ only because we imply the energy portion $\hbar\omega$ transported by the wave, the portion being present in the resonance condition.

Let us demonstrate first that, in the case relevant to the Compton effect, the denominators in (4.1) cannot vanish. In order to employ formulae of the type (4.1) we assume that the electron is placed in a large box, so that its energy spectrum is discrete. At the same time, if the volume of the box tends to infinity, the energy will be practically continuous. Let the coordinate system be chosen such that the electron is at rest before the scattering. In this case the initial energy of the electron is $E_m = m_0 c^2$ while the final energy is $E_n = mc^2$ where $m_0$ is the rest mass of the electron and $m = m_0/(1 - v^2/c^2)^{1/2}$ with $v$ being the velocity of the electron after the scattering. Since $E_n > E_m$ and $\omega > 0$, the second denominator in (4.1) cannot vanish. Equating the first denominator to zero yields $\hbar\omega = (m - m_0)c^2$. The energy $\hbar\omega$ transmitted to the electron entails a transmitted momentum of $\hbar\omega/c$ according to the foregoing. However, it is readily seen that the equation $\hbar\omega/c = mv$ is incompatible with the preceding equation $\hbar\omega = (m - m_0)c^2$. Therefore, the first denominator in (4.1) cannot vanish either in the present case. It should be observed that in our argumentation we are compelled to resort to supplementary considerations other than an equation of the type (4.1) because, when deducing (4.1), the quantum system is implied to be immobile as distinct from the free electron.

Seeing that the first approximation gives nothing, we are coming to a second approximation of the perturbation theory noting in passing that the Compton effect in quantum electrodynamics is a second-order process as well [7]. If use is made of Landau and Lifshitz's [10], Sec. 40, notation, the equation for a second-order correction to $a_n$ is of the form

$$i\hbar \frac{da_n^{(2)}}{dt} = \sum_k V_{nk}(t) a_k^{(1)} . \qquad (5.4)$$

In the case of the Compton effect, we must take into account that the electron is affected not only by the incident wave with the frequency $\omega$ but also by the scattered wave with a frequency $\omega'$. For this reason, instead of being given by Landau and Lifshitz's equation (40.6), the perturbation operator should be taken to be

$$\mathsf{V} = \mathsf{F}\, e^{-i\omega t} + \mathsf{G}\, e^{i\omega t} + \mathsf{F}'\, e^{-i\omega' t} + \mathsf{G}'\, e^{i\omega' t} . \qquad (5.5)$$

The first approximation $a_n^{(1)}$ will now be given by a formula similar to (4.1) to which must be added analogous terms with $\omega'$ and $F'_{nm}$ replacing $\omega$ and $F_{nm}$ respectively.

Upon placing this $a_n^{(1)}$ in (5.4) it is not difficult to solve the resulting equation. We shall not write down the full expression for $a_n^{(2)}$ restricting ourselves only to terms that can describe a resonance. With use made of the same notation as in (4.1), we have

$$a_n^{(2)} = \frac{1}{\hbar^2 (\omega_{nm} - \omega + \omega')} e^{i(\omega_{nm} - \omega + \omega')t} \sum_k \left( \frac{F_{km} F'^*_{kn}}{\omega_{km} - \omega} + \frac{F_{nk} F'^*_{mk}}{\omega_{km} + \omega'} \right). \qquad (5.6)$$

It is interesting to observe that the sum in (5.6) contains terms of two types just as in quantum electrodynamics in which two Feynman diagrams correspond to the Compton effect [7]. Only the denominator of the first factor can vanish here if one recalls the discussion of the first approximation. If the denominator is set equal to zero, one obtains the resonance condition

$$\hbar\omega + E_m = \hbar\omega' + E_n .\tag{5.7}$$

Substituting the above expressions for $E_m$ and $E_n$ yields

$$\hbar\omega + m_0 c^2 = \hbar\omega' + mc^2 .\tag{5.8}$$

To this must be added the law of momentum conservation. Upon recalling the relation between the energy and momentum of an electromagnetic wave indicated above and the fact that the electron was at rest before the scattering, we can match each term in (5.8) with the corresponding momentum, so that

$$\hbar\mathbf{k} = \hbar\mathbf{k}' + m\mathbf{v}.\tag{5.9}$$

Equations (5.8) and (5.9) are identical with the well-known relations obtained usually for the Compton effect on the basis of the corpuscular view on light (see, e.g., Blokhintsev [13] who, by the way, contends that these relations can by no means be understood in terms of the undulatory view on light). As to the angular distribution of the scattered radiation, Klein and Nishina [21] derived their famous formula without using the concept of a particle-like photon, all the more so as in the nonrelativistic limit the Klein-Nishina formula goes over into Thomson's formula obtained in classical electrodynamics [7,14].

It should be remarked that in accord with experiment [5] there must be negligible delay in the Compton effect because the resonance can occur only if the scattered wave with the frequency $\omega'$ is present simultaneously with the incident wave in view of (5.5). Besides, on physical grounds it is apparent that the electron must begin to emit radiation practically at the instance the incident wave hits the electron setting it into vibration and thus imparting an acceleration to it (see $\tau_{rd} \sim 10^{-23}$ s that follows from (3.1)). One further remark needs to be made. The manifestation of the Compton effect may well be affected by the nonuniformities of the incident electromagnetic wave mentioned above. Owing to the nonuniformities the electron may recoil in different directions with respect to the propagation direction of the wave. Explanation of this fact from the undulatory point of view presented difficulties [4].

It is worthy of note that (5.6) can be applied not only to free electrons but also to systems that possess a discrete spectrum, e.g., to atoms, molecules, crystals. In this case, (5.7) is relevant to Raman scattering [7,13] which consequently is a resonance phenomenon as well.

A wave explanation of the Compton effect was proposed by Schrödinger [22] as far back as 1927 but it is seldom if ever for that paper to be cited nowadays. Schrödinger reasoned in parallel with Brillouin who had earlier on considered reflection of a light wave at a sound

423wave. In this context there appeared a relation analogous with the well-known Bragg formula for X-ray diffraction in a crystal. In place of the sound wave, Schrödinger took the de Broglie electronic wave and, on a base of the relation, obtained a condition equivalent to (5.8) and (5.9) (in a four-dimensional notation). Note that in our explanation the electron is regarded as a particle (more precisely, our approach does not require to take the wave properties of the electron into explicit account), whereas in Schrödinger's explanation the electron as well as light is a wave. It should be added that the Bragg formula is in fact a resonance condition.

Having considered the photoelectric and Compton effects from the undulatory point of view we now turn to other aspects of the problem under discussion. Once there are no particle-like photons, the question arises as to what a 'photon counter' shows. The photon counter registers events of resonance interaction of the incident electromagnetic wave with atoms of the working substance of the counter. Analogously with (5.2), the number of these events per unit time and thereby the count rate are proportional to the intensity of the incident light.

One can, in a like manner, explain gradual formation of the picture on a photographic plate (or another detector) in the case of a faint light flux, e.g., formation of a diffraction pattern. One observes first randomly located separate points that subsequently build up into the picture. When light is absorbed by molecules of the photoemulsion, different processes take place. The absorption will be efficacious and will give energy necessary for the processes only if the molecules come into resonance with the incident light. Although the photographic plate is exposed to a continuous electromagnetic wave, the resonance interaction between the molecules and light happens at random at different points of the photoemulsion where favourable conditions for the resonance come about at a given instant of time. Therefore, the point-made picture attests the atomic structure of matter rather than the particle-like character of light.

It is instructive to discuss two known experiments that were regarded as impressive evidence for the corpuscular nature of electromagnetic radiation. In the experiment by Joffé and Dobronrawov [23], a small particle of dust suspended in an electric field was exposed to a very faint flux of X-rays. Every 30 minutes on the average a photoelectric event occurred causing the particle to wince because of emission of an electron, the wince being observed visually. When discussing this experiment, one asserts that the experiment can be explained only if one invokes the concept of a particle-like photon that gives energy to the electron instantaneously, otherwise it is quite incomprehensible why only one out of a wealth of electrons in the particle accumulates energy during 30 minutes in order to be able to escape from the particle when the energy suffices for this. If one leans upon the above presented undulatory point of view, does no electron in the particle accumulate energy during half an hour. On the average every half-hour there occurs resonance interaction of one of the electrons with the incident electromagnetic wave (the probability of the interaction being proportional to the intensity of the wave according to (5.2) was very low in that experiment). The electron sucks in the whole of the energy almost instantaneously (more precisely, with the speed of light) and escapes from the particle. All of this looks as if a particle has suddenly struck the electron and knocked it out of the metal.

Another experiment, by Bothe [24], was in essence as follows. A metallic foil was exposed to X-rays and emitted secondary X-rays owing to X-ray fluorescence. The foil was placed between two X-ray counters, so that the counters received the X-rays from the opposite sides of the foil. If atoms of the foil emitted spherical X-ray waves, the counters should come into action simultaneously. In the experiment, however, the counters gave readings incoherently with each other. One concluded from this that the X-rays were made up of particles emitted chaotically in various directions.

When explaining that experiment in undulatory terms, one must take account of two circumstances. From the above discussion of the function of a 'photon counter' it follows that the point in time of the resonance interaction between the incident radiation and an atom of the working substance of the counter is determined by processes that go on in the counter itself, and does not depend upon the processes in another counter. For this reason the readings of the counters, though excited by one and the same wave, must not necessarily correlate with each other. The second circumstance is due to the nonuniformities of actual electromagnetic waves mentioned above. An atom need not emit a spherically symmetric wave because when radiating energy the atom passes into a nonstationary state, whereas such states are not, in general, spherically symmetric (see, e.g. Eq. (2.22)). Consequently, the radiation should be direction-dependent. Besides, the radiation can reflect at other atoms of the foil. As a result, when escaping from the foil, the radiation due to a certain atom may predominantly follow only one direction, so that one of the counters will receive most of the radiation and register it, while the other getting a minor portion will not come in action because the probability of a count event depends upon the radiation intensity.

We end this section by remarking that, when deriving his formula for blackbody radiation, Planck assumed only that an oscillator can emit or absorb energy in discrete amounts (quanta) equal to $\hbar\omega$ but he did not imply that the electromagnetic radiation was made up of particles. Planck's formula is relevant to a state of thermodynamic equilibrium, that is, to the most probable state of a system. We saw in Sec. 4 that the energy is most efficaciously emitted or absorbed in the form of packets $|E_n - E_m|$ while, on the other hand, $|E_n - E_m| = \hbar\omega$. Even the radiation that is due to the transition from a nonstationary state to a stationary one and that is not emitted in packets can in a later time be absorbed in a resonance way and reemitted as a portion of $|E_n - E_m| = \hbar\omega$; besides, this radiation is not dominant. Seeing that the above packets play a leading role in the processes in question one can, when considering the thermodynamic equilibrium, proceed from a simplifying assumption that the oscillators of which the system is made up exchange energy solely in quanta of $\hbar\omega$, and so deduce the Planck formula. In this connection it may be noted that the relation $\hbar\omega = |E_n - E_m|$ for the frequency emitted or absorbed is simplified even from the viewpoint of the photonic concept because spectral lines have a non-zero natural breadth (cf. Introduction). For this reason, in reality the Bohr relation $\hbar\omega = |E_n - E_m|$ determines the most probable frequency alone, and here again we encounter only the most probable values as above.

## 6. Modern experiments on properties of light

The last decades are marked by a rich variety of experiments performed as to properties of light. Sometimes authors assert that their experiments attest the corpuscular nature of light much more convincingly than the experiments discussed in the preceding section. Amongst the new experiments one may set out two main groups.

The first group comprises experiments on the test of Bell's inequalities. Let us recall in brief the essence of these inequalities. Let there be an apparatus that measures some characteristic of an object (a particle), e.g., a component of the spin of an electron or the polarization of a 'photon'. In classical physics the characteristic had, prior to measuring, a definite, though unknown, value. Bell [25] has shown that this permits one to deduce an inequality, say $A \geq 0$, where the expression for $A$ depends upon the characteristic measured and conditions of the experiment. As to quantum mechanics, prior to measuring the characteristic had no definite value and such an inequality is lacking. Violation of a Bell inequality in an experiment suggests that properties of the object under study cannot be described in classical terms (without special assumptions such as non-locality of interaction) and thereby the object reveals quantum properties.

Experimental tests of the Bell inequalities were performed with light as well in order to clarify properties of photons (see, e.g., [26,27]; a comprehensive review of such experiments can be found in [28]). One observed violation of Bell's inequalities in these experiments, from which one inferred that the photons behaved according to quantum mechanical laws like other atomic particles. At the same time it is necessary to remark that experimental verification of Bell's inequalities is not a simple matter, and serious doubt was thrown as to the interpretation of the experiments [29,30].

We now turn to discussing the Bell inequalities from the point of view presented in this paper. According to this viewpoint photons are nonexistent in nature. This assertion is even stronger than in quantum mechanics: prior to measuring, not only is a characteristic of the object lacking but the object itself is also lacking. An imitation of the object (the 'photon') appears only during the measurement, which is due to resonance interaction of the light wave with the apparatus. Therefore, as in quantum mechanics, one is to observe violation of Bell's inequalities here. In this connection we may quote Belinskii [31] who remarks that optical experiments can be interpreted as demonstrating the absence of photons till the registration of light by a detector.

The idea of the second group of the experiments mentioned at the outset of this section is as follows. A light beam is directed onto a beamsplitter that breaks down the beam into two beams. One of them which traverses the beamsplitter without deflection is incident on a detector $D_t$ while the other that is reflected by the beamsplitter perpendicularly to the initial beam hits another detector $D_r$. The authors of the idea reason along the following lines. Let the initial light be so feeble that it contains only one photon at a given interval of time. If the photon is a particle, it must be indivisible. Therefore the photon either will traverse the beamsplitter in a straight line and will be registered by the detector $D_t$ or will be reflected completely, in which case it will be registered by the detector $D_r$. This being so, the count events of the detectors must not coincide in time if the experiment is carried out with a number of separate photons. The experiments, in principle, corroborated the absence of coincidences [32,33], which proved in the authors' opinion the corpuscular nature of light more convincingly than the photoelectric effect.

Such experiments can however be explained on the basis of the above undulatory standpoint as well. When light falls on a beamsplitter, an atom or a group of atoms (for short, we shall speak of an atom) in the beamsplitter can come into resonance interaction with the incident light and thus suck in an energy portion of $E_n - E_m$. It should be emphasized that some atoms of the beamsplitter do must interact strongly with light as long as the beamsplitter is not fully transparent. Returning to the normal state the atom will emit an ordinary wave train whose intensity will be far above the intensity of the incident faint light. For this reason the wave train will be readily absorbed and reemitted by other atoms of the beamsplitter. These processes will not permit the train to spread out. As a result, according to optical laws the train either will emerge from the beamsplitter in the direction of $D_t$ or will be reflected towards $D_r$. Consequently, the detectors will not give simultaneous counts. It should be added that an experimental setup often contains additional equipment such as different optical plates, filters etc in which the resonance interaction of the light wave with atoms can occur as well. For this reason too, the input and output light in the setup may differ drastically, the latter being made up of intense wave trains going randomly in the mutually perpendicular directions even if the former is a continuous plane wave of weak intensity.

However, as was mentioned in the preceding section, there are no ideal plane waves in nature and actual waves are nonuniform, the nonuniformities being most conspicuous in the case of a faint light flux. This entails another mechanism that can explain the results of the above experiments even if the resonance interactions are not taken into account. The half-silvered mirror of the beamsplitter is not homogeneous on the microscopic level. The beamsplitter transmits 50% of light and reflects 50% of light only if the intensity of light is sufficiently high when the light flux can be considered to be spatially uniform. The picture can cardinally change if the light flux is weak and essentially nonuniform. When one states that there is only one 'photon' in an experimental device, this in fact amounts to saying that there is a wave train of finite length emitted by only one atom in the source of light. The wave train can hit such a site in the mirror of the beamsplitter that will allow the train to pass almost completely in the direction of $D_t$. On the other hand, the wave train can hit a site where the train is reflected almost completely, so that it will actuate $D_r$. In order to establish the possibility of this course of events, one ought to know the microscopic structure of the beamsplitter and the processes that cause transmission or reflection of light in it.

The results of the above experiments can also be influenced by the fact that processes which lead to registration of light in the detectors do not correlate between the detectors as was noted when discussing the Bothe experiment in the preceding section. Even if one and the same light wave falls on the detectors, the count events in the detectors do not need to coincide in time, which will be most pronounced in the case of a faint light flux.

We shall not discuss details of the experimental data presented in [32,33] since, if one starts from the above undulatory view on light, it is clear that the result of an experiment of this type depends upon many factors such as the resonance interactions that may occur or may be lacking, the character of the wave train emitted by the source, the microscopic structure of the beamsplitter and so on. These factors important in the case of a faint light flux are not taken into account in the simplified equations used in [32,33]. The equations are simplified because all devices employed are regarded as structureless objects with uniform properties while dealing with such delicate things as a separate wave train emitted by a single atom. Factually, the experimental data presented in [32,33] attest an atomic structure of the devices rather than the particle-like character of light analogously with the gradual formation of a diffraction pattern discussed in the preceding section. It should be mentioned that there is another point of view as to why the beamsplitter experiment [32,33] does not conclusively confirm the particle behaviour of light [34].

Summing up the consideration of the preceding and present sections we can see that all effects and phenomena that seemingly confirm the corpuscular nature of light can be explained on the basis of the undulatory theory of light. The traditional explanation of these effects and phenomena for which it is sufficient to resort solely to the concept of a photon suggests a unique mechanism that lies at the root of these. The proposed undulatory explanation is based on a unique mechanism as well, which is resonance interaction of the electromagnetic wave with the quantum system. For this reason the explanation of an effect that is not mentioned in the present paper can readily be transformed from photonic terms into undulatory terms upon changing some phrases slightly and invoking the idea of the resonance. For example, various 'photon' experiments in which statistical regularities are studied can be regarded as experiments on statistics of resonance interactions of light with matter.

At the same time, the second mechanism considered just above can play a role as well, namely, the wave train radiated by an atom or a nucleus can imitate a photon. Probably, it is just this mechanism that is essential in the case of $\gamma$-quanta that represent electromagnetic wave trains of high energy. The behaviour of such a bundle of energy may resemble the behaviour of a particle inasmuch as diffraction is scarcely pronounced once the wavelength is small. Besides, there exists such a phenomenon as self-focusing known from nonlinear optics. Self-focusing can occur when a powerful electromagnetic wave propagates in a medium whose dielectric permeability $\varepsilon$ depends upon the wave intensity [35]. For a vacuum, $\varepsilon \equiv 1$. The wave, however, produces virtual electron-positron pairs because of which the vacuum becomes polarized and behaves as a medium with wave-intensity dependent $\varepsilon$ [7], Sec. 129. Owing to self-focusing the $\gamma$-bundle of energy can even contract when moving.

## 7. Concluding remarks

The present paper shows that, when trying to understand the principles of quantum mechanics, account must be taken of the fact that the objects of investigation for quantum mechanics are charged particles, systems of charged particles or particles creating magnetic fields. Only if the electromagnetic radiation that always accompanies nonstationary processes in the above objects is properly allowed for, does the peculiar role played by stationary states become evident. In the paper we emphasized also the role of resonance phenomena in which the stationary states manifest themselves once again. Although the resonance phenomena are well-known, their important role in quantum mechanics was not assessed at its true worth.

In the paper it was demonstrated in particular that the resonance phenomena can explain various effects that are ascribed usually to manifestation of particle-like photons. Consequently, arguments that are traditionally put forward to confirm the existence of the photons are in no way justified, and therefore light as well as any other electromagnetic radiation possess undulatory properties alone. It is to be emphasized that this in no wise excludes the existence of dual (corpuscular and undulatory) properties as to genuine particles such as an electron or a nucleon.

As mentioned in Introduction, the photon is one of the basic notions in quantum electrodynamics. At the same time, the photon in quantum electrodynamics is not at all a tiny ball; the term 'photon' implies an elementary excitation of the electromagnetic field in the form of a plane wave with energy $\hbar\omega$ (see Introduction). Let us discuss the question as to why this notion is helpful in quantum electrodynamics. In quantum electrodynamics one starts from an initial state of a system and finds out the final state without considering the time development of relevant processes; various observables are calculated as well. The observables are often represented in terms of the traces of some operators. The trace does not depend upon the choice of functions with whose help it is calculated provided they constitute a complete orthonormal set. As shown in this paper, energy is most efficaciously absorbed and emitted by a quantum system only in packets equal to $|E_n - E_m| = \hbar\omega$. In quantum electrodynamics, this is taken into account in advance by breaking the initial electromagnetic field up into elementary plane waves, each of which having an energy of $\hbar\omega$. On the other hand, $\hbar\omega$ should be equal to $|E_n - E_m|$. To take this into account as well, one endows the elementary plane waves with properties of excitations of a harmonic oscillator whose energy eigenvalues are equal to $E_n$, and thus introduces creation and annihilation operators that add or subtract the elementary wave excitations (this procedure is clearly described by Bogoliubov and Shirkov [9], Sec. 6; see also [36]). If one employs these excitations, the traces will be expanded in powers of $\alpha$ and can be calculated sequentially. Therein lies the meaning of the quantization of the electromagnetic field in quantum electrodynamics. Thus we see that the notion of the photon in quantum electrodynamics is merely a convenient mathematical tool that enables one to consider the result of a process and calculate observables without solving time-dependent equations describing the process.

Quantum electrodynamics studies also processes referred to as photon-photon interactions. As long as the quantum electrodynamical photon is a wave, the photon-photon interaction is in fact an interaction of electromagnetic waves. The Maxwell equations are linear and cannot describe this interaction. The electromagnetic waves, however, polarize the vacuum (see the end of Sec. 6) and interact via the vacuum polarization. In this connection let us discuss Feynman diagrams. A Feynman diagram is a schematic and pictorial representation of an involved mathematical formula that contains electromagnetic potentials and electron-positron wavefunctions. There is nothing peculiar to a particle-like photon in the formula. Consequently, the photonic lines in the diagram should be regarded as lines relevant to electromagnetic waves. It will be recalled that the electromagnetic energy portion $\hbar\omega$ possesses a momentum $\mathbf{p} = \hbar\mathbf{k}$ ascribed customarily to a photon (see the discussion of the Compton effect in Sec. 5).

Of interest is to discuss the process termed the creation of an electron-positron pair by a photon. The energy of the pair is $2mc^2$. Replacing $E_n - E_m$ in (4.3) by $2mc^2$ yields

$$\omega = \frac{2mc^2}{\hbar}. \tag{7.1}$$

Customarily this equation is written as $\hbar\omega = 2mc^2$ where $\hbar\omega$ is called the energy of the photon. In actual fact, (7.1) is a resonance condition. If the frequency of an electromagnetic wave coincides with (7.1), there occurs resonance. The energy of the wave is efficaciously sucked in the region where the resonance happens (cf. Sec. 5). When the energy sucked in attains $2mc^2$, the pair will be created. The annihilation of the pair is described by the second term of (4.1) in line with the last but one paragraph of 4. As a result, the pair will be converted into electromagnetic waves whose frequency is given by (7.1) again.

It is instructive to cite the example of another 'particle' akin to the photon, namely, the phonon. Any solid contains only atoms, molecules, electrons, and there are no particles such as phonons in the solid. The phonon is an excitation (a wave) concerning all constituents of the solid. Upon breaking the wave field up into excitations with energy $\hbar\omega$ one may count the excitations and apply Bose statistics to them. The concept of the phonons is helpful in studies of low-temperature properties of the solid when the number of excitations is small, for example, when deducing the Debye law for the heat capacity. At the same time, the law can be obtained without using the concept of the phonon [37]. By the way, high-temperature properties of the same solid that are described by classical physics are explained without invoking the concept.

It is necessary also to underline another important point without which it is impossible to explain, on undulatory grounds, the effects ascribed usually to particle-like photons. The effects can be explained only if account is taken of the reverse impact of the quantum system on the incident wave and the resulting essential distortion of the wave in case a resonance occurs. This feedback was overlooked in previous attempts to understand the effects physically. The reason is that the standard perturbation theory employed for most of calculations in quantum mechanics is developed only on the assumption that the perturbation is fixed, whereas this assumption holds solely in the absence of the resonance. Besides, in quantum mechanics and especially in quantum electrodynamics it is common practice, as mentioned above, to consider only the initial and final states when the wave and the system do not interact. In this case the incident wave is not distorted by the resonator, of course.

**References**

[1]  M. O. Scully and M. Sargent III, Physics Today **25**, 38 (1972).

[2]  R. Kidd, J. Ardini, and A. Anton, Amer. J. Phys. **57**, 27 (1989).

[3]  W. E. Lamb Jr., Appl. Phys. B **60**, 77 (1995).

[4]  J. N. Dodd, J. Phys. B: Atom. Molec. Phys. **8**, 157 (1975).

[5]  R. Kidd, J. Ardini, and A. Anton, Amer. J. Phys. **53**, 641 (1985).

[6]  P. N. Kaloyerou, Phys. Rep. **244**, 287 (1994).

[7]  V. B. Berestetskii, E. M. Lifshitz, and L. P. Pitaevskii, Quantum Electrodynamics (Pergamon, Oxford, 1982).

[8] S. S. Schweber, An Introduction to Relativistic Quantum Field Theory (Row and Peterson, Evanston, 1961).

[9] N. N. Bogoliubov and D. V. Shirkov, Quantum Fields (Benjamin/Cummings, London, 1983).

[10] L. D. Landau and E. M. Lifshitz, Quantum Mechanics (Pergamon, Oxford, 2000).

[11] L. I. Schiff, Quantum Mechanics (McGraw-Hill, New York, 1968).

[12] V. P. Bykov, Usp. Fiz. Nauk **161** (10), 145 (1991) [Sov. Phys. Usp. **34,** 910 (1991)].

[13] D. I. Blokhintsev, Quantum Mechanics (Reidel, Dordrecht, 1964).

[14] L. D. Landau and E. M. Lifshitz, The Classical Theory of Fields (Pergamon, Oxford, 1994).

[15] V. S. Krivitskii and V. N. Tsytovich, Usp. Fiz. Nauk **161** (3), 125 (1991) [*Sov. Phys. Usp.* **34,** 250 (1991)].

[16] J. A. Stratton, Electromagnetic Theory (McGraw-Hill, New York, 1941).

[17] J. Franck and G. Hertz, Verhand. Deut. Physik. Ges. **16**, 457, (1914).

[18] A. I. Akhiezer and V. B. Berestetskii, Quantum Electrodynamics (Interscience Publ., New York, 2004).

[19] J. W. S. Rayleigh, The Theory of Sound, vol. 2 (Dover, New York, 1976).

[20] E. Schrödinger, Brit. J. Philos. Sci. **3**, 109, 233 (1952).

[21] O. Klein and Y. Nishina, Zs. f. Phys. **52,** 853 (1929).

[22] E. Schrödinger, Ann. Physik **387** (**82**), 257 (1927).

[23] A. Joffé and N. Dobronrawov, Zs. f. Phys. **34**, 889 (1925).

[24] W. Bothe, Zs. f. Phys. **37,** 547 (1926).

[25] J. S. Bell, Physics (NY) **1**, 195 (1964).

[26] A. Aspect, P. Grangier, and G. Roger, Phys. Rev. Lett. **47**, 460 (1981).

[27] Y. H. Shih and C. O. Alley, Phys. Rev. Lett. **61**, 2921 (1988).

[28] A. V. Belinskii and D. N. Klyshko, Usp. Fiz. Nauk **163** (8), 1 (1993) [Phys. Usp. **36**, 653 (1993)].

[29] E. Santos, Phys. Rev. Lett. **66**, 1388 (1991).

[30] L. De Caro and A. Garuccio, Phys. Rev. A **50**, R2803 (1994).

[31] A. V. Belinskii, Usp. Fiz. Nauk **173**, 905 (2003) [*Phys. Usp.* **46**, 877 (2003)].

[32] J. F. Clauser, Phys. Rev. D **9**, 853 (1974).

[33] A. Aspect, P. Grangier, and G. Roger, J. Optics (Paris) **20**, 119 (1989).

[34] P. N. Kaloyerou, J. Phys. A: Math. Gen. **39**, 11541 (2006).

[35] L. D. Landau and E. M. Lifshitz, Electrodynamics of continuous media (Pergamon, Oxford, 1984).

[36] J. M. Ziman, Elements of Advanced Quantum Theory (University Press, Cambridge, 1969).

[37] V. A. Golovko, Physica A **310**, 39 (2002).